\documentclass[pra,twocolumn,aps,showpacs,floatfix,groupedaddress,10pt]{revtex4-1}

\usepackage{graphicx}
\usepackage{rotating}\bf 
\usepackage{amsmath}
\usepackage{bbm}
\usepackage{subfigure}
\usepackage{color}
\usepackage[utf8]{inputenc}

\renewcommand{\v}[1]{{\bf #1}}

\newcommand{\be}{\begin{equation}}
\newcommand{\ee}{\end{equation}}
\newcommand{\bea}{\begin{eqnarray}}
\newcommand{\eea}{\end{eqnarray}}
\newcommand{{\br}}{\bf r}
\newcommand{{\up}}{\uparrow}

\begin{document}

\title{Structure of the first order Reduced Density Matrix in three electron systems: A Generalized Pauli Constraints assisted study}

\author{Iris Theophilou}
\affiliation{Max Planck Institute for the Structure and Dynamics of Matter and Center for Free Electron Laser Science, 
22761, Hamburg, Germany}
\author{Nektarios N.\ Lathiotakis}

\affiliation{Theoretical and Physical Chemistry Institute, National Hellenic 
Research Foundation, Vass.\  Constantinou 48, GR-11635 Athens, Greece}
\author{Nicole Helbig}
\affiliation{Peter-Gr\"unberg Institut and Institute for Advanced Simulation,
Forschungszentrum J\"ulich, D-52425 J\"ulich, Germany}

\begin{abstract}
We investigate the structure of the one-body Reduced Density Matrix (1RDM) of three electron systems, i.e.\ doublet and quadruplet spin configurations, corresponding to the smallest interacting system with an open-shell ground state. To this end, we use Configuration Interaction (CI) expansions of the exact wave function in Slater determinants built from natural orbitals in a finite dimensional Hilbert space. With the exception of maximally polarized systems, the natural orbitals of spin eigenstates are generally spin dependent, i.e.\ the spatial parts of the up and down natural orbitals form two different sets. A measure to quantify this spin dependence is introduced and it is shown that it varies by several orders of magnitude depending on the system. We also study the ordering issue of the spin-dependent occupation numbers which has practical implications in Reduced Density Matrix Functional Theory minimization schemes when Generalized Pauli Constraints (GPCs) are imposed and in the form of the CI expansion in terms of the natural orbitals. Finally, we discuss the aforementioned CI expansion when there are GPCs that are almost "pinned".
\end{abstract}

\maketitle

\section{Introduction}

Reduced density matrices are used in different approximative methods for solving the many-body problem. For example, the one-body reduced density matrix (1RDM) serves as the basic variable in Reduced Density Matrix Functional Theory (RDMFT) \cite{Review-RDMFT_Katarzyna_Klaas,MULLER1984,BBC3,Solids,Bench,PNOF7,Bloch2017}, where the total energy is approximated by a functional of the 1RDM. With the inclusion of fractional occupation numbers, RDMFT has the potential to improve the description of strongly correlated  systems \cite{mott_jctc,pina} that are very difficult to describe within density functional or Hartree Fock theories. In order to ensure that the ground-state 1RDM, obtained from a RDMFT minimization scheme, corresponds to fermions, the so-called ensemble $N$-representability conditions \cite{Coleman} are enforced as constraints on the occupation numbers. Recently, the 1RDM has also be used in a geometric method to understand noise-assisted energy transfer \cite{Mazziotti_Chakraborty-energy-transfer-1RDM,Mazziotti-Chakraborty-GPC-excited}. Instead of the 1RDM, one can employ the two-body reduced density matrix (2RDM) as the basic variable, in which case the energy functional is known exactly \cite{mazziotti_review}. Although the complete set of $N$-representability conditions can be constructed for the 2RDM \cite{Nrep-2RDM}, in practice, only a limited number of them are implemented \cite{Mazziotti_variation_optimization}. Consequently, employing either the 1RDM or the 2RDM as the basic variable, one obtains approximate density matrices by the minimization procedure. Recently, a systematic derivation of the generalized Pauli constraints (GPCs) for a given number of electrons in a finite dimensional Hilbert space was presented in Ref.~\cite{Klyachko_GPC}. The GPCs ensure that the 1RDM is not only fermionic but corresponds to a pure state rather than an ensemble. Hereafter, it was shown that the approximate density matrices, obtained within RDMFT or a 2RDM approach, suffer from some deficiencies which can be removed if the energy minimization is constrained so that the 1RDM satisfies the GPCs \cite{GPC_RDMFT, DePrinceGPC}.

In electronic structure theory, one often focuses on reproducing the physical behavior of systems with an even number of particles that form closed-shell systems, i.e.\ spin singlets. Open-shell systems are usually treated by extending the closed-shell approximations or even just applying them without any modifications. For example, most RDMFT functionals, although devised for closed-shells, are extended to treat open-shell systems \cite{Open-Shell-RDMFT-Nic-Nek,Open-Shell-RDMFT-Pernal}. With only a small number of 
exceptions \cite{Open-Shell-RDMFT-Pernal,Solids}, 
these generalizations and the corresponding calculations typically use spin-independent natural orbitals, i.e.\ the eigenfunctions of the 1RDM are treated to be  the same in both spin channels while the occupation numbers can differ. While this treatment is correct for spin singlets and states with maximum total spin $S$, generally open-shell systems with a different total spin cannot be treated accurately this way \cite{triplet}.

Already decades ago, the natural orbitals have been considered as the ideal single-particle basis for the convergence of a CI expansion in Slater determinants to the many-body wavefunction \cite{L1955,Coleman}. Although it was later found that this claim is only true for two-particle systems \cite{CI_NOs_Klaas,CI_NOs_not_perfect}, the natural orbitals are still a good basis for the fast convergence of a CI expansion. For example, for the fully polarized linear equidistant H$_3$ system in equilibrium geometry, we find that the 12 natural orbitals with the largest occupation numbers are sufficient to construct a CI wave function which yields the correct total energy within the chemical accuracy (1kcal/mol), while 45 Hartree-Fock orbitals are needed to achieve the same accuracy. In the past, little attention was paid to this type of expansion since the exact natural orbitals can only be obtained by solving the interacting many-body problem. In light of the recently derived GPCs, the expansion of the exact many-body wave function in terms of Slater determinants built from natural orbitals has gained increasing attention, as "pinning" of a GPC leads to zero expansion coefficients for certain Slater determinants \cite{Christian_PhysRevLett,Christian-Carlos-proof_selection_rule,Extension_HF}. A recent study shows that using the simplified CI ansatz that is implied by the pinning of a GPC, in cases where the corresponding GPC is only almost pinned, yields wave functions that can recover a significant part of the correlation energy \cite{Mazziotti_Chakraborty-sparsity}.

In order to truly allow for the application of density matrix methods to open-shell systems, it is crucial to understand the structure of the 1RDM beyond closed-shell configurations. To this end, we study a few different three-electron systems which can form spin doublets or spin quadruplets. We calculate the exact wave function in a restricted active space and obtain the corresponding 1RDM, its natural orbitals and occupation numbers. The natural orbitals are used as a single-particle basis to reconstruct the exact wave function. For specific examples we can show that in the doublet case the natural orbitals have to be spin dependent, i.e.\ the spatial parts of up and down orbitals do not coincide. The GPCs are always given as conditions on  ordered occupation numbers. However, these occupation numbers correspond to spin up or down natural orbitals in all states except for the fully polarized ones. We discuss the possible orderings of spin occupation numbers for three electrons in Hilbert spaces of dimension 6 or 8. For the larger Hilbert space, the sets of GPCs will be different for different orderings of the occupation numbers which has practical implications since we find different orderings present in different systems. 

The paper is structured as follows: In Section \ref{sec:method} we provide a short overview of the GPCs, the theoretical background for this work, and the numerical details of our investigation. In Section \ref{sec:spin_dependence}, we discuss the spin-dependence of the natural orbitals by studying three electron systems with total spin $S=1/2$ or $S=3/2$.  Note that we always consider Hamiltonians that commute with $\mathbf{S^2}$, thus $S$ is a good quantum number. In Section \ref{sec:ordering}, we discuss the problem of the ordering of the occupation numbers when spin is taken into account, in relation to GPCs, which is relevant in any open-shell system. Finally, in Section \ref{sec:quasipinning}, we discuss the relation between a GPC being almost pinned and the coefficients in the CI expansion in terms of natural orbitals using one of the systems we studied and a Hilbert space with dimension 8. We conclude our findings in section \ref{sec:conclusions}.

\section{The Method}\label{sec:method}
\subsection{Generalized Pauli constraints}
The generalized Pauli constraints ensure that a given 1RDM not only represents a fermionic ensemble but actually corresponds to a fermionic wave function. For closed-shell systems with time-reversal symmetry, one does not need to consider the GPCs, as the necessary and sufficient condition for a 1RDM to be pure-state $N$-representable is that it satisfies the ensemble $N$-representability conditions and, in addition, its occupation numbers, in the two spin channels are identical \cite{S1966}. For open-shell systems or systems without time-reversal symmetry, the GPCs can be derived for a given number of particles $N$ and a given size of the Hilbert space of the natural orbitals $M$ \cite{Klyachko_GPC}. 
The number of constraints increases quickly with both $N$ and $M$, and, for practical applications, one considers typically only small values for $N$ and $M$. Furthermore, the GPCs, for three electron systems that are considered here, have been derived in practice only for $M\leq 12$ \cite{ChristianarXiV}. We emphasize that the size of the 
Hilbert space refers to the number of natural orbitals that are assumed to have non-zero occupation number, while the basis set in which we expand these natural orbitals can be much larger than $M$. 

The GPCs are linear constraints on the occupation numbers $n_j$ and have the form 
\begin{equation}
\kappa_0+\sum_{j=1}^M \kappa_j n_j\geq 0\,,
\label{eq:gpc}
\end{equation}
with integer coefficients $\kappa_j$. Note that the occupation numbers are ordered in non-increasing order, i.e.\ $n_1\geq n_2\geq n_3 ...$. Special consideration needs to be given to cases where a constraint is saturated, i.e.\ it is satisfied as an equality rather than an inequality and, in addition, there are no degenerate occupation numbers. In this situation, a configuration interaction (CI) expansion of the many-body wave function in terms of Slater determinants built from the natural orbitals is simplified since the constraint removes certain determinants from the expansion \cite{Christian-Carlos-proof_selection_rule}. However, the constraints which are satisfied as equalities, if there are any, depend strongly on the particular system \cite{Mazziotti_Chakraborty-GPC1,Pinning_Tennie_Christian}. There is one particular case, namely that with $N=3$ and $M=6$, in which three out of the four constraints are actually given as equalities. The constraints, in this case, were initially derived by Borland and Dennis \cite{Borland-Dennis} in the 70's while their necessity was proven by Ruskai just a decade ago \cite{Ruskai}. 
The constraints in this case read 
\begin{eqnarray}
\nonumber
n_1+n_6=1,\quad && n_2+n_5=1,\\
n_3+n_4=1,\quad && n_5+n_6-n_4\geq 0.
\end{eqnarray}
One can write the equality constraints in operator form  and apply these operators to a CI expansion with Slater determinants expressed in terms of natural orbitals. Thus, e.g.\ $n_1+n_6-1=0$ corresponds to the operator $\hat{n}_1+\hat{n}_6-\hat{1}$. In this case, each Slater determinant that contains either orbital $\phi_1$  or $\phi_6$ is an eigenstate of this operator with zero eigenvalue. For determinants that contain neither orbital $\phi_1$ nor $\phi_6$ the eigenvalue is $-1$, whereas, for Slater determinants that contain both orbitals the eigenvalue is $+1$. One can show that only those determinants that have a zero eigenvalue for all constraint operators can be part of the CI expansion \cite{Christian-Carlos-proof_selection_rule}. Consequently, the CI expansion for three electrons and $M=6$ reads as
\begin{eqnarray}
\nonumber
|\Psi\rangle&=&c_{1}|123\rangle+c_2|124\rangle+c_3|135\rangle+c_{4}|145\rangle\\
&&+c_5|236\rangle+c_6|246\rangle+c_{7}|356\rangle+c_8|456\rangle,
\label{eq:Psi_Borland-Dennis}
\end{eqnarray}
i.e.\ it consists of eight different Slater determinants. With $|ijk\rangle$ we denote a Slater determinant that contains the natural orbitals $\phi_i$, $\phi_j$, $\phi_k$. As discussed various times in the  literature, see for example \cite{Carlos_satur_36}, if the inequality is satisfied as an equality, another five determinants are removed leaving a total wave function containing only the determinants $|123\rangle$, $|145\rangle$, and $|246\rangle$.

\subsection{Numerical details}
\label{subsec:numerical}

In order to illustrate our findings we perform Complete Active Space Self Consistent Field Calculations (CASSCF), using the GAMESS computer code \cite{gamess}, for a set of three electron systems in spaces of dimension 6, 8 and 9 using an extended basis set. We extract the exact natural orbitals in these spaces and thus we can discuss either the CI expansion in these natural orbitals or the corresponding GPCs. Note here that the GPCs are different for different dimensions of the Hilbert space. 

We investigate different three electron systems, namely the Li atom, the LiH$^+$ molecule at equilibrium and dissociation geometry, an equilateral H$_3$ triangle and three different linear H$_3$ configurations. For the linear H$_3$ systems we choose two equidistant configurations, one at equilibrium geometry at bond length 0.9~\AA\  and one at 0.7~\AA . We also choose a non-equidistant linear H$_3$ configuration with bond lengths 0.5~\AA~and 1.3~\AA . The bond length of the triangular hydrogen system is 0.9~\AA . The natural orbitals are expanded in Gaussian basis sets. For Li we use the aug-cc-pCVQZ basis set while for H$_3$ we employ the aug-cc-pVQZ basis set \cite{basis_set}. In all calculations, the systems have spin $S=S_z=1/2$ and the natural orbital spaces have dimension $6$ or $8$. The only exception is the calculation of the number of natural orbitals necessary to obtain the total energy within chemical accuracy, which was carried out for the $S=S_z=3/2$ state in a Hilbert space of dimension 9.

\section{Spin dependence of natural orbitals}\label{sec:spin_dependence}

In this section, we discuss under which conditions the spatial parts of up and down natural orbitals are the same, in the case of Hilbert spaces with dimension 6 and 8. If the wave function in terms of natural orbitals describes a maximally polarized quadruplet, no special considerations about the spin need to be taken, since all electrons are of the same spin and one can, therefore, ignore it and consider only the spatial parts. However, in the case that three-electron systems form doublets, one needs to take into account the spin dependence of the spatial parts of the natural orbitals.

\subsection{Natural orbitals of three electron doublets in a Hilbert space of dimension 6}

Performing a CASSCF calculation for the set of systems described in Section \ref{subsec:numerical} with a doublet spin configuration, in a restricted Hilbert space with three natural orbitals per spin, we always find the following ordering of the six occupation numbers 
\begin{equation}\label{eq:ordering3-6}
n_{1\uparrow}\geq n_{2\uparrow}\geq n_{1\downarrow}\geq n_{3\uparrow}\geq n_{2\downarrow}\geq n_{3\downarrow}\,,
\end{equation}
where spin-up is assumed to be the majority spin channel.
As we will discuss later, this ordering is unique apart from an interchange of $n_{1\downarrow}$ and $n_{3\uparrow}$. However, this interchange does not affect the three equality GPCs which now read
\begin{align}
\nonumber
n_{1\uparrow}+n_{3\downarrow}&=1,& 
n_{2\uparrow}+n_{2\downarrow}=1,\\
\label{eq:GPCs3-6spin}
n_{1\downarrow}+n_{3\uparrow}&=1.
\end{align}
For the ordering (\ref{eq:ordering3-6}), the inequality condition $n_{2\downarrow}+n_{3\downarrow}-n_{3\uparrow}\geq 0$ is always satisfied as an equality. Since the sum of occupation numbers with spin down is fixed to one, it reduces to the condition $n_{1\downarrow}+n_{3\uparrow}\leq 1$ which is one of the equality constraints above. As we discuss in Appendix \ref{app:3-6GPC}, only three Slater determinants contribute to the total wave function 
\begin{equation}\label{eq:psi_spin_half}
|\Psi\rangle=c_{121}|1^\uparrow 2^\uparrow 1^{'\downarrow}\rangle+c_{132}|1^\uparrow 3^\uparrow 2^{'\downarrow}\rangle +c_{233}|2^\uparrow 3^\uparrow 3^{'\downarrow}\rangle.
\end{equation}
Note that the ordering \eqref{eq:ordering3-6} and the resulting wavefunction \eqref{eq:psi_spin_half} were discussed in \cite{Carlos-3_6}, as it was found to be the only possible ordering compatible with the two largest occupations belonging to the majority spin channel.
Let us now make the assumption that the natural orbitals are spin independent, i.e.\ the spatial parts of orbitals $i^\uparrow$, $i^{'\downarrow}$, for $i=1,2,3$, are identical. Then, the first and the last of the determinants in (\ref{eq:psi_spin_half}) are eigenstates of the total spin $\hat{\v S}^2$ with $S=1/2$. The second determinant, however, cannot be an eigenstate of $\hat{S}^2$ unless the spatial parts of the natural orbitals are spin dependent. For a different assumption, namely the following pairs of orbitals with the same spatial part, $(1^\uparrow|1^{'\downarrow})=
\int d^3r \: \varphi_{1\uparrow}^*(\v r)\varphi{'}_{1\downarrow}(\v r)=1$, $(2^\uparrow|3^{'\downarrow})=1$, and $(3^\uparrow|2^{'\downarrow})=1$, all three determinants in (\ref{eq:psi_spin_half}) are eigenstates of $\hat{\v S}^2$  with $S=1/2$. However, in an atomic system like the Li atom this assignment implies that the second largest occupation number in one spin channel is of $s$-character while in the other spin channel it has $p$-character. Under normal conditions, this appears to be very unlikely. A similar argument can be used for the assignment $(1^\uparrow|2^{'\downarrow})=1$, $(2^\uparrow|1^{'\downarrow})=1$, and $(3^\uparrow|3^{'\downarrow})=1$. Here, the largest occupation number in one spin channel would be the 1s orbital while in the other spin channel it would be the 2s orbital. If one of the coefficients in Eq.\ (\ref{eq:psi_spin_half}) is zero, there are additional possibilities, for example, if $c_{233}=0$ the assignment $(1^\uparrow|3^{'\downarrow})=1$, $(2^\uparrow|1^{'\downarrow})=1$, and $(3^\uparrow|2^{'\downarrow})=1$ yields a spin eigenstate. All these cases represent explicit exceptions that are not realized in general. Hence, from the wave function (\ref{eq:psi_spin_half}), we conclude that in general the spatial parts of natural orbitals have to be spin dependent. This implies that none of the Slater determinants is an eigenstate of the total spin by itself anymore. In order for the wave function to represent an eigenstate of the total spin with $S=1/2$, the spin contaminations from the three different determinants have to cancel out. Consequently, the natural orbitals corresponding to the two spin channels, although different from each other, have to span the same space. Therefore, one can expand the natural orbitals of a spin channel in terms of the orbitals of the opposite one (see also Appendix \ref{app:eigenstate_cond} for an explicit example). 

Furthermore, the three coefficients $c_{ijk}$ have to satisfy certain relations necessary for the wave function to represent a spin eigenstate. Since the orbitals are different in the two spin channels, these conditions are nontrivial and depend on the overlaps between the orbitals in the two channels. As a simple example, we consider the case where the spatial parts of the orbitals $2^\uparrow$ and $2^{'\downarrow}$ are orthogonal to each other, i.e.\ $(2^\uparrow|2^{'\downarrow})=0$.
Since the wavefunction $\Psi$ is an eigenstate of $\hat{\v S}^2$, the remaining orbitals have to satisfy the condition
\begin{equation}
c_{121}(3^\uparrow|1'^{\downarrow})=-c_{233}(1^\uparrow|3'^{\downarrow}).
\end{equation}
A more detailed discussion of the relation between the orbitals of the two spin channels and the conditions between the coefficients of the expansion is given in Appendix~\ref{app:eigenstate_cond}.

\subsection{Quantifying the difference between spin up and spin down natural orbitals}

In the above discussion, we demonstrated analytically that the spatial parts of spin up and spin down natural orbitals are in general different, in the case of a Hilbert space of dimension 6. This ``spin dependence'' is expected in general for any dimension of the Hilbert space for systems that are not in either singlet or in spin configuration with maximal $S$. We introduce the following definition
\begin{equation}
\Delta_\mathrm{spin}=1-\frac{1}{M}\sum_{j=1}^M \max_{k=1,M}\biggl|\int d^3r \varphi_{j\uparrow}^*(\v r)\varphi'_{k\downarrow}(\v r)\biggl|
\end{equation}
which quantifies the deviation of the spatial parts of the natural orbitals in the up and down spin channels. Thus, for identical natural orbitals in both spin channels $\Delta_\mathrm{spin}=0$.

We calculated this quantity for three electron systems, in restricted Hilbert spaces of dimension 6 and 8 using the natural orbitals obtained from CASSCF calculations in the corresponding spaces.
\begin{table}
\begin{tabular}{l|cc}
& \multicolumn{2}{c}{$\Delta_\mathrm{spin}$}\\
System & $\:\:M=6\:\:$ & $\:\:M=8\:\:$\\ \hline
Li & 0.003 & 0.003\\
LiH$^+$ equilibrium & 0.003 & 0.002 \\
LiH$^+$ dissociation & 0.000 & 0.000\\
H$_3$ triangular & 0.063 & 0.115 \\
H$_3$ linear, equidistant, small distance \ & 0.000 & 0.130\\
H$_3$ linear, equidistant, equilibrium &  0.005 & 0.004\\
H$_3$ linear, non-equidistant & 0.075 & 0.097 \\
\end{tabular}
\caption{\label{tab:deltas} 
Difference between the exact natural orbitals of the two spin channels for different systems using 6 or 8 natural orbitals.}
\end{table}
In Table \ref{tab:deltas}, the quantity $\Delta_\mathrm{spin}$ for different systems is presented. As one can see, the only case where for both $M=6$ and $M=8$ the set of spin up and spin down natural orbitals are identical is the LiH$^+$ at dissociation. This is a special case with a configuration where there is a pair of electrons, one with spin up and one with spin down, localized at the Lithium cation and a third one localized at the hydrogen atom. Consequently, the system consists of two subsystems, a closed-shell lithium cation and a fully polarized hydrogen atom, both of which can be described with spin-independent natural orbitals (see section \ref{sec:maxpol} for a detailed analysis of maximally polarized systems). For the wave function, we find that $c_{132}=0$ in Eq.\ (\ref{eq:psi_spin_half}) and there are two pairs of natural orbitals, i.e.\ $(1^\uparrow|1'^\downarrow)=1$ and $(3^\uparrow|3'^\downarrow)=1$. Therefore, both determinants which contribute to the wave function are eigenstates of the total spin. The behavior of $\Delta_\mathrm{spin}$ for the linear equidistant H$_3$ at the small distance is somewhat unexpected. 
Apparently for $M=6$ the orbitals are identical in the two spin channels while for $M=8$ we find that they are strongly ``spin dependent''. A detailed investigation of the wave function for the $M=6$ case shows that only two determinants contribute to the wave function, the coefficient $c_{233}$ in Eq.\ (\ref{eq:psi_spin_half}) being zero. 
As a consequence, the occupation number $n_{3\downarrow}$ is zero. We find again two pairs of natural orbitals with identical spatial parts, $(2^\uparrow|1'^\downarrow)=1$ and $(3^\uparrow|2'^\downarrow)=1$. However, in this case, contrary to that of LiH$^+$ in the dissociation limit, there is no explanation based on geometry for the zero value coefficients in (\ref{eq:psi_spin_half}). Due to our findings for $M=8$ we conclude that the space with only six natural orbitals is too limited for a correct representation of the natural orbitals of this system. For the remaining systems, it is not clear if the spin dependence of the orbitals increases or decreases with $M$. Most likely, there are two competing effects: the variational freedom increases on going from $M=6$ to $M=8$, allowing thus weaker ``spin dependence". The ``spin dependence" of the $M=6$ case was due to the fact that only three of the nine possible Slater determinants in the CI expansion can have non-zero coefficients. On the contrary, the $M=6$ case can be very restrictive, as we have seen for the linear H$_3$ discussed earlier, which implies that the only way to construct a spin eigenstate is to set one of the coefficients in Eq.\ (\ref{eq:psi_spin_half}) equal to zero. In this case, by increasing the variational freedom the ``spin dependence" increases.

\subsection{Natural orbitals of states with maximum $\v S^2$}
\label{sec:maxpol}

For three electrons, the maximum possible total spin is $S=3/2$ with the four possible values for the $z$-component $S_z=-3/2, -1/2, 1/2, 3/2$.
For the maximally polarized state $|S=3/2, S_z=3/2\rangle$, there is no question of spin-dependence of the spatial parts of the natural orbitals since only the up channel has non-zero occupations. Therefore, we can choose any set of orbitals for the down channel without changing the 1RDM, since they have zero occupations. In a restricted space of only three natural orbitals for each spin channel, the wave function of the maximally polarized state is a single Slater determinant,
\be
|S=3/2, S_z=3/2\rangle=|1^\uparrow2^\uparrow3^\uparrow\rangle\,.
\ee
The other states of the quadruplet configuration can be obtained by applying successively the $\mathbf{\hat{S}_-}$ operator to this state flipping one spin at a time to the down direction. For example, we get
\bea\nonumber
\lefteqn{
|S=3/2, S_z=1/2\rangle=}\hspace{1cm}\\
\label{S=1.5S=0.5}
&&\frac{1}{\sqrt{3}}\left(|1^\downarrow2^\uparrow3^\uparrow\rangle+|1^\uparrow2^\downarrow3^\uparrow\rangle+|1^\uparrow2^\uparrow3^\downarrow\rangle\right).
\eea
Since $\mathbf{\hat{S}_-}$ only affects the spin degrees of freedom, the spatial parts of the natural orbitals for the two spin channels are identical. Applying the $\mathbf{\hat{S}_-}$ operator again, the same arguments can be used for the natural orbitals of the $|S=3/2, S_z=-1/2\rangle$ and $|S=3/2, S_z=-3/2\rangle$ states. Therefore, for all the states with maximum total spin the natural orbitals can be chosen to be spin-independent. Although, in this example, we used three electrons and three spatially orthogonal natural orbitals in each spin channel, the same arguments hold generally for states with maximal total spin $S$ irrespective of the number of electrons and the number of natural orbitals which are used to expand the corresponding wave function.

\section{Ordering of the occupation numbers}\label{sec:ordering}

\subsection{Practical implications of different orderings of spin-dependent occupation numbers}
So far we have discussed the spin dependence of the natural orbitals. Additionally, the occupation numbers are also spin dependent.
If we fix the number of spin up $N^\uparrow$ and spin down $N^\downarrow$ electrons the ensemble $N$-representability  conditions \cite{Coleman} must be satisfied per spin channel i.e.
\begin{equation}
\sum n_{i\uparrow}  = N^\uparrow\nonumber\,,\qquad
\sum n_{i\downarrow} = N^\downarrow \,.
\end{equation}
The above conditions are typically used as constraints in RDMFT minimizations when the ground state of odd-particle systems is calculated. 

As we discussed already, the GPCs (\ref{eq:gpc}) are additional conditions that involve the occupation numbers of the 1RDM so that it corresponds to a pure state. Contrary to the ensemble $N$-representability conditions, these conditions assume the occupation numbers indexed in non-increasing order, i.e.\ $n_1\geq n_2\geq n_3 ...$. This ordering ignores the spin index. If both spin channels have non-zero occupation numbers we can order them for each spin channel separately, $n_{1\sigma}\geq n_{2\sigma}\geq n_{3\sigma} ...$. 
However, one does not know in general how a specific occupation number from one spin channel compares to the occupation numbers in the other spin channel. As an example  we consider three electron systems with two up and one down electron. For not too strongly correlated systems, we expect two occupation numbers in the up channel and one in the down channel close to one and all remaining occupation numbers close to zero. We can therefore choose the two large occupation numbers in the up channel as $n_{1\uparrow}$ and $n_{2\uparrow}$ with $n_{1\uparrow}\geq n_{2\uparrow}$ and $n_{1\downarrow}$ as the largest occupation number in the down channel. Obviously, we do not know a priori if $n_{1\downarrow}$ is smaller or larger than the two large occupation numbers in the up channel. 

The ordering of spin indexed occupation numbers is essential for many approximate functionals in RMDFT, e.g.\ for possible open-shell extensions of functionals that separate the orbitals according to their occupation \cite{BBC3}. Also, as we will demonstrate for the case of $M=8$, there is not a unique way to express GPCs in terms of the spin indexed occupation numbers. 
This means that if we choose to employ these conditions as constraints in an actual RDMFT minimization we might need to employ a different expression of constraints in every iteration. For the exact functional, this issue does not exist since one only needs to use the ensemble $N$-representability conditions, making thus the GPCs redundant \cite{V1980}. However, for approximate RDMFT functionals, the GPCs are violated when the ensemble conditions alone are employed during an actual minimization \cite{GPC_RDMFT}. 

The non-uniqueness of ordering is also relevant for the CI expansion of the wave function in terms of natural orbitals. If there is a pinned GPC, i.e.\ a GPC which is satisfied as an equality, as discussed earlier, certain Slater determinants from the CI expansion will be removed. It is worth to mention that, the same pinned GPC written in descending order occupation numbers, ignoring spin, would remove different determinants from the expansion depending on the ordering of occupation numbers. 

\subsection{ Orderings of occupation numbers for doublets in spaces of 6 or 8 orbitals}

As discussed above, the existence of different orderings of the spin-dependent occupation numbers in different systems leads to complications in practical applications of the GPCs. In this section, we examine whether 
different orderings of spin-dependent occupation numbers appear in practice
for systems of three electrons in Hilbert spaces of dimensions 6 and 8.
We find that the ordering indeed depends on the system. For example, for the Li atom and $M$=8, we find the following ordering 
\begin{equation}\label{eq:orderingLi}
n_{1\uparrow} > n_{2\uparrow} = n_{1\downarrow} > n_{3\uparrow} = n_{2\downarrow} = n_{3\downarrow} = n_{4\uparrow} > n_{4\downarrow}\,,
\end{equation}
while for the LiH$^+$ molecule we find
\begin{equation}
n_{1\uparrow} > n_{1\downarrow} \geq n_{2\uparrow} > n_{3\uparrow} > n_{2\downarrow} >  n_{4\uparrow} > n_{3\downarrow} > n_{4\downarrow}.
\end{equation}

All orderings of occupation numbers for the systems studied here, fall in three groups. In group 1, the ordering is
\begin{equation}\label{eq:case1}
n_{1\uparrow} \geq n_{2\uparrow} \geq n_{1\downarrow} \geq n_{3\uparrow} \geq n_{2\downarrow} \geq n_{3\downarrow} \geq n_{4\uparrow} \geq n_{4\downarrow}\,,
\end{equation}
while, in group 2, 
\begin{equation}\label{eq:case2}
n_{1\uparrow} \geq n_{2\uparrow} \geq n_{1\downarrow} \geq n_{3\uparrow} \geq n_{2\downarrow} \geq n_{4\uparrow} \geq n_{3\downarrow} \geq n_{4\downarrow},
\end{equation}
i.e.\ the occupation numbers $n_{3\downarrow}$ and $n_{4\uparrow}$ are exchanged compared to the ordering of group 1. Finally, in group 3,
\begin{equation}\label{eq:case3}
n_{1\uparrow} \geq n_{1\downarrow} \geq n_{2\uparrow} \geq n_{3\uparrow} \geq n_{2\downarrow} \geq n_{4\uparrow} \geq n_{3\downarrow} \geq n_{4\downarrow}.
\end{equation}
The occupation numbers $n_{2\uparrow}$ and $n_{1\downarrow}$ in the ordering of the group 3 have exchanged their positions in relation to that of group 2. As we see, in ordering groups 
\eqref{eq:case1}, \eqref{eq:case2}, \eqref{eq:case3}, neither the ordering of the large occupation numbers nor that of the small ones remains the same in all groups.

The ordering of the occupation numbers for
the linear non-equidistant H$_3$ and the linear equidistant H$_3$ at equilibrium geometry are 
in group 1.
 For the linear equidistant H$_3$ at a smaller than the equilibrium distance, the ordering falls into the second group while for LiH$^+$ at equilibrium geometry into the third. The remaining three systems, the Li atom, the LiH$^+$ at dissociation distance and the equilateral H$_3$, contain degenerate occupation numbers. Therefore, the ordering of the occupation numbers is not unique. Comparing Eq.\ (\ref{eq:orderingLi}) with Eqs.(\ref{eq:case1}) - (\ref{eq:case3}) shows that any of the three orderings can be assigned to the Li atom. The same is true for the LiH$^+$ at dissociation. In the case of equilateral H$_3$, the ordering of the occupation numbers falls into group 2, however, $n_{3\uparrow}$ and $n_{2\downarrow}$ are degenerate.

Since the ordering of the occupation numbers is indeed different for different three electron systems, the question arises whether the set of GPCs are truly different. Of course, most GPCs are different if the ordering changes but one could imagine a situation where one constraint switches its role with another one. To clarify this question, we compare the constraints for the first two cases explicitly. (The full lists of constraints for all three cases are given in Appendix \ref{app:3-8GPC}). As these two cases differ only in an exchange of ordering affecting the occupation numbers $n_{3\downarrow}$ and $n_{4\uparrow}$, it is clear that any constraint which does not include either of these two occupation numbers is identical in both cases. This is true for several constraints (3, 5, 9, 11, 17, 19 and 21). There are also three constraints (12, 13 and 24) which contain the sum $n_{3\downarrow}+n_{4\uparrow}$ and are therefore also unaffected by the change in order. Then, we have two conditions (15 and 16 in case 1) which are interchanged (16 and 15 in case 2). Without these obviously identical constraints there remain 19 constraints which appear to be different in the two cases. We tried to identify other constraints or combinations of them that are the same for the two cases but could not find any beyond the ones mentioned above. We also found that if the first constraint in case 2 is satisfied as an equality and there are no degenerate occupation numbers, 12 Slater determinants from the CI expansion of the many-body wave function would be removed. We could not find any combination of constraints in case 1 which causes removal of the same set of determinants. Hence, we conclude that the GPCs take indeed a different form in these two cases and suspect that the same is true for any two different orderings of the occupation numbers.\par

In the case of three electrons in six natural orbitals three in each spin channel, there are only two possibilities for the ordering of the occupations. The one we always encountered numerically from our CASSCF calculations in 3-6
\begin{equation}
i)\quad  n_{1\uparrow}\geq n_{2\uparrow}\geq n_{1\downarrow}\geq n_{3\uparrow}\geq n_{2\downarrow}\geq n_{3\downarrow}
\end{equation}
and the following one
\begin{equation}
ii)\quad  n_{1\uparrow}\geq n_{2\uparrow}\geq n_{3\uparrow}\geq n_{1\downarrow}\geq n_{2\downarrow}\geq n_{3\downarrow}\,.
\end{equation}
As already discussed in section \ref{sec:spin_dependence}, the inequality GPC for the first case is satisfied as an equality. For the second ordering, where $n_{1\downarrow}$ and $n_{3\uparrow}$ are interchanged, the inequality reads as
\begin{eqnarray}
n_{2\downarrow}+n_{3\downarrow}-n_{1\downarrow}\geq 0\,,
\end{eqnarray}
which implies that $n_{1\downarrow}\leq 0.5$. However, it is rather unlikely, in practice, for the biggest spin-down occupation to have such a small value and that explains why we find only the first ordering in our calculations. Note that in the set of systems that we explored numerically the lowest value for $n_{1\downarrow}$ is $0.98$. Whether the inequality is pinned or not has no effect on the wave function form in terms of natural orbitals which always reads
\begin{equation}\label{eq:psi_spin_half_2ndtime}
|\Psi\rangle=c_{121}|1^\uparrow 2^\uparrow 1^{'\downarrow}\rangle+c_{132}|1^\uparrow 3^\uparrow 2^{'\downarrow}\rangle +c_{233}|2^\uparrow 3^\uparrow 3^{'\downarrow}\rangle.
\end{equation}
Of course, the wave function might contain even less determinants if some of the coefficients $c_{ijk}$ are zero.

As we show in Appendix \ref{app:3-6GPC}, all the other possible orderings one could construct lead to degenerate occupation numbers and could be considered as special cases of the aforementioned orderings. Both orderings lead to the same three equality GPCs (see Eq.\ (\ref{eq:GPCs3-6spin}) and Appendix \ref{app:3-6GPC}).


\section{Effect of ``Quasipinning" on the structure of the wave function}
\label{sec:quasipinning}
So far, we have discussed the effect of a truly pinned GPC on the structure of the wave function. The question arises whether a constraint that is almost pinned (quasipinned) leads to certain Slater determinants with a very small
coefficient. Those determinants are those that the constraint would remove if it was fully pinned. This would imply that there is an ansatz of the wave function in terms of natural orbitals where one knows a priori that there is a number of Slater determinants which contribute with small coefficient and can then be excluded from the variational optimization \cite{Christian-Carlos-proof_selection_rule,Extension_HF}. 

As a test case we present the results for the linear H$_3$ at equilibrium geometry, in a doublet state, in the $M=8$ space. In this example, we do not have degenerate occupation numbers and one can apply the pinned GPCs as operators to the wave function \cite{Christian-Carlos-proof_selection_rule}. In this system, there is no GPC which is exactly pinned. There are three constraints which are very close to the border, namely numbers 5, 8 and 12 in Table \ref{tab:case1}. The left-hand-sides of the inequalities are $1.9\cdot 10^{-11}$, $1.7\cdot 10^{-9}$, and $1.8\cdot 10^{-9}$, respectively. The next smallest value we find is $1.7 \cdot 10^{-3}$. Looking at the coefficients in the CI expansion, we find four determinants with coefficients larger than $4.0 \cdot 10^{-2}$ while the remaining coefficients are of the order of $10^{-5}$ or smaller. In other words, in the constraints there is a clear distinction between those that are close to the border and those that are not. Similarly, for the CI coefficients we can clearly distinguish between small and large coefficients. We now assume that all three constraints that are very close to the border are actually at the border, i.e.\ that they are satisfied as equalities. We then find that only 6 Slater determinants remain in the CI expansion. The Slater determinants with the four largest CI coefficients are among these i.e. they have a zero eigenvalue for all three constraints. However, already the determinant with the fifth largest coefficient should be removed by the second and third constraints. There are two Slater determinants that are allowed by the constraints, however, their coefficients are very small, presumably for other reasons than GPCs. 

We verified that
all three constraints, satisfy the recently proven inequality\cite{Christian-Carlos-proof_selection_rule, Quasipinning_Schilling} (see Eq. (15) of Ref.\cite{Christian-Carlos-proof_selection_rule}) that relates the structure of the wave function to quasipinning. In other words, for each considered quasipinned  constraint, the distance of the constraints from being pinned multiplied by two is larger than the sum of the squares of the coefficients of the determinants that are removed from the wave function in case of full pinning. Thus, our findings  support the claim that this inequality is a useful tool to analyze the structure of the wave function.

\section{Conclusions}\label{sec:conclusions}
In this paper, we studied the 1RDM of three electron systems that form doublets or quadruplets in restricted Hilbert spaces. In a 6-dimensional natural orbital space, we demonstrated that, for the doublet, the spatial parts of the up and the down natural orbitals form different sets. We provided a theoretical explanation based on the CI expansion in terms of the natural orbitals. We also introduced a measure to quantify the deviation between up and down spatial parts and we found by studying three electron systems that ``spin dependence"  is largely system dependent. This result is important for the RDMFT minimization, where typically one set of spatial orbitals for both up and down spin is used. While this is justified for quadruplet states of three-electron systems, and in general for states with maximal total spin $S$, we have seen that the ``spin dependence" can be significant in other cases. In other words, for an accurate description of general open-shell systems one needs to extend the currently used RDMFT implementations and allow for ``spin-dependent" natural orbitals.\par

We also studied the possible ordering of occupation numbers in both spin channels for a natural orbital space of dimension 8 using the same set of three electron systems forming doublets. We found different orderings realized in different systems which then lead to different expressions of the Generalized Pauli Constraints. This has implications whenever one employs the GPCs as constraints during a RDMFT minimization since one would have to deal with different expressions of GPCs not only for different systems but also during the minimization procedure. The GPCs have gathered some interest over the last years for studying the general properties of the 1RDM in relation to the CI expansion of the many-body wave function in terms of natural orbitals. The possible different orderings of occupation numbers of different spin channels has also implications for the simplifications in the CI expansion. 

Finally, we used a specific example of the numerical CI expansion in terms of natural orbitals to explore the relation between GPCs being almost pinned, known as quasipinning, and coefficients of particular Slater determinants being close to zero. We found that quasipinning is consistent with the structure of the wave function at a quantitative level since a recently introduced measure was found to be satisfied.

\appendix

\section{Generalized Pauli Constraints for different spin ordering}\label{app:GPC}
\subsection{$N=3, M=6$}\label{app:3-6GPC}
As discussed in section \ref{sec:spin_dependence}, numerically we find only one ordering of the occupation numbers. Since different orderings are in principle possible for this case, the question remains why they are not realized in any of the systems that are considered here. In this Appendix we consider all orderings that are possible and the implications the corresponding GPCs have on the occupation numbers and, as a consequence, on the structure of the wave function. As usual, the occupation numbers are ordered in non-increasing order separately for each spin channel. 

We consider a wave function with $S_z=1/2$, therefore, each Slater determinant contributing to the wave function, contains two orbitals for spin up and one for spin down. We divide the $M=6$ orbitals into two sets, three orbitals per spin channel. The number of possible Slater determinants is then nine and, if there were no additional constraints, the wave function  would be written as
\bea\nonumber
\lefteqn{|S=1/2, S_z=1/2\rangle=}\hspace{1cm}\\
\nonumber
&& c_{121} |1^\uparrow2^\uparrow1'^\downarrow\rangle
+c_{122} |1^\uparrow2^\uparrow2'^\downarrow\rangle
+c_{123} |1^\uparrow2^\uparrow3'^\downarrow\rangle\\
\nonumber
&+&c_{131} |1^\uparrow3^\uparrow1'^\downarrow\rangle
+c_{132} |1^\uparrow3^\uparrow2'^\downarrow\rangle
+c_{133} |1^\uparrow3^\uparrow3'^\downarrow\rangle\\
\nonumber
&+&c_{231} |2^\uparrow3^\uparrow1'^\downarrow\rangle
+c_{232} |2^\uparrow3^\uparrow2'^\downarrow\rangle
+c_{233} |2^\uparrow3^\uparrow3'^\downarrow\rangle,\\
\label{eq:doublet}
\eea
where the coefficients $c_{jkl}$ are labeled according to the orbitals which appear in the determinant. The prime on the orbitals of the down channel indicates that the orbitals can differ in the two spin channels. We note that for the above wave function to consist of natural orbitals additional constraints between the expansion coefficients $c_{jkl}$ must be satisfied to make the constructed 1RDM diagonal. However, as we discuss in the following, for most orderings the GPCs impose even stricter constraints on the wave function form, which automatically results in a diagonal 1RDM. There are two orderings where the off-diagonal elements appear and one needs to impose additional conditions on the involved coefficients so that they become zero (see cases 2a ii) and iii) below).

From the number of electrons in the up channel being two, we can conclude that the two largest occupation numbers cannot both have spin down and the two smallest occupation numbers cannot both have spin up. Also, due to the separate ordering in the two spin channels, the largest occupation number overall is either $n_{1\uparrow}$ or $n_{1\downarrow}$ while the smallest occupation number is either $n_{3\uparrow}$ or $n_{3\downarrow}$. We now consider each case separately and study the implications of all four GPCs starting with the three equalities.

\begin{itemize}
\item[Case 1:] $n_{1\downarrow}$ is the largest occupation number.

It follows immediately that $n_{1\uparrow}$ is the second largest occupation number. We now investigate the two possibilities for the smallest occupation number
\begin{itemize}
\item[a)] $n_{3\downarrow}$ is the smallest occupation number.

The first GPC reads $n_{1\downarrow}+n_{3\downarrow}=1$. As the sum of down occupations is one it follows that $n_{2\downarrow}=0$ which implies that $n_{3\downarrow}=0$ because of the ordering, and consequently $n_{1\downarrow}=1$. Hence, in Eq.\ (\ref{eq:doublet}) only $c_{121}$, $c_{131}$, and $c_{231}$ can be non-zero. The complete ordering for this case is given by
\bea\nonumber
n_{1\downarrow}\geq n_{1\uparrow}\geq n_{2\uparrow}\geq n_{3\uparrow}\geq n_{2\downarrow}\geq n_{3\downarrow}.
\eea
The inequality GPC, therefore, reads as $n_{2\downarrow}+n_{3\downarrow}-n_{3\uparrow}\geq 0$. As the two down occupation numbers are zero, this can only be satisfied if $n_{3\uparrow}=0$ from which we immediately conclude that $n_{1\uparrow}=n_{2\uparrow}=1$ as the sum of the up occupations is two. Hence, only the coefficient $c_{121}$ can be non-zero and the wave function is a single Slater determinant.

\item[b)] $n_{3\uparrow}$ is the smallest occupation number.

It follows that the second smallest occupation number is $n_{3\downarrow}$, since, as we mentioned before, the two smallest occupations cannot both belong to spin up. Hence, the three equality GPCs read as
\begin{align}\nonumber
n_{1\downarrow}+n_{3\uparrow}&=1, \: n_{1\uparrow}+n_{3\downarrow}=1\\
n_{2\uparrow}+n_{2\downarrow}&=1.
\end{align}
Acting with these three GPCs on the wave function (\ref{eq:doublet}) we see that only the coefficients $c_{121}$, $c_{132}$, and $c_{233}$ can be non-zero. Then, the two largest and the two smallest occupation numbers are given in terms of the coefficients as
\begin{eqnarray}
n_{1\downarrow}=|c_{121}|^2,& &n_{1\uparrow}=|c_{121}|^2+|c_{132}|^2,\\
n_{3\downarrow}=|c_{233}|^2,& &n_{3\uparrow}=|c_{132}|^2+|c_{233}|^2.
\end{eqnarray}
As we required $n_{1\downarrow}\geq n_{1\uparrow}$ it follows that $c_{132}=0$ and then, due to $n_{3\downarrow}\geq n_{3\uparrow}$, $c_{233}=0$. Again, only $c_{121}$ remains and the wave function is a single Slater determinant. At this point, one might argue that the removal of Slater determinants from the wave function due to the equality GPCs was not appropriate since we are having degenerate occupation numbers in the end. Hence, we assume a degeneracy with either $n_{1\downarrow}=n_{1\uparrow}$ or $n_{1\uparrow}=n_{2\uparrow}$ from the beginning and investigate the effect on the coefficients in Eq.\ (\ref{eq:psi_spin_half_2ndtime}). We only need to consider the degeneracies in the three largest occupation numbers since the degeneracies in the small occupation numbers follow directly from the three equality GPCs. Also, the degeneracy $n_{1\downarrow}=n_{2\uparrow}$ implies that, due to the ordering, the largest three occupation numbers are all degenerate. It can therefore be regarded as a special case of the two cases mentioned above. For both degeneracies one can  show that the only possibility to satisfy the equality GPCs and the ordering is for all coefficients but $c_{121}$ to vanish. Hence, one arrives at the same single Slater determinant as with the above procedure which would have allowed for non-degenerate occupation numbers initially.

Hereafter, the only possible ordering that we find starting from the hypothesis that $n_{1\downarrow}$ is the largest occupation number is
\begin{eqnarray}
n_{1\downarrow}&=&n_{1\uparrow}= n_{2\uparrow}=1,\nonumber\\
n_{3\uparrow}&=&n_{2\downarrow}=n_{3\downarrow}=0.
\end{eqnarray}
The state consists of only one Slater determinant,
\bea
|S=1/2, S_z=1/2\rangle=
 |1^\uparrow2^\uparrow1'^\downarrow\rangle,
\eea
and the inequality constraint is pinned as all of the occupations involved are zero. The aforementioned ordering can be seen as a special case of the ordering that we find numerically (\ref{eq:doublet}).
\end{itemize}
\item[Case 2:] $n_{1\uparrow}$ is the largest occupation number.
\begin{itemize}
\item[a)] $n_{3\downarrow}$ is the smallest occupation number.
 
The second largest occupation number can be either $n_{1\downarrow}$ or $n_{2\uparrow}$ while the second smallest occupation number is either $n_{3\uparrow}$ or $n_{2\downarrow}$. Therefore, we find the following five possibilities for the overall ordering of the occupation numbers
\begin{eqnarray}
i) &  n_{1\uparrow}\geq n_{1\downarrow}\geq n_{2\uparrow}\geq n_{2\downarrow}\geq n_{3\uparrow}\geq n_{3\downarrow}\\
ii) &  n_{1\uparrow}\geq n_{1\downarrow}\geq n_{2\uparrow}\geq n_{3\uparrow}\geq n_{2\downarrow}\geq n_{3\downarrow}\\
iii) &  n_{1\uparrow}\geq n_{2\uparrow}\geq n_{1\downarrow}\geq n_{2\downarrow}\geq n_{3\uparrow}\geq n_{3\downarrow}\\
iv) &  n_{1\uparrow}\geq n_{2\uparrow}\geq n_{1\downarrow}\geq n_{3\uparrow}\geq n_{2\downarrow}\geq n_{3\downarrow}\\
v) &  n_{1\uparrow}\geq n_{2\uparrow}\geq n_{3\uparrow}\geq n_{1\downarrow}\geq n_{2\downarrow}\geq n_{3\downarrow}
\end{eqnarray}
The first equality reads as $n_{1\uparrow}+n_{3\downarrow}=1$ for all cases.
For the cases ii) and iii) we find
\begin{equation}
n_{1\downarrow}+n_{2\downarrow}=1, \: n_{2\uparrow}+n_{3\uparrow}=1
\end{equation}
We can then immediately conclude that $n_{3\downarrow}=0$ and $n_{1\uparrow}=1$ due to the sum of occupation numbers per spin channel. Therefore, only the coefficients $c_{121}$, $c_{131}$, $c_{122}$, and $c_{132}$ can be non-zero since all the determinants that contain the $3'^\downarrow$ orbital have to vanish and the same is true for the determinants that do not contain $1^\uparrow$. As the smallest occupation is zero the inequality constraint can be satisfied only if the fifth largest occupation number is equal to or larger than the forth largest occupation number. This is only possible if the fourth and the fifth occupation in descending order are equal. This means for both cases ii) and iii) that $n_{3\uparrow}=n_{2\downarrow}$ which implies for the coefficients that $|c_{131}|=|c_{122}|$, which results in $n_{2\uparrow}=n_{1\downarrow}$. Thus, both orderings ii) and iii) reduce to the following one:
\begin{eqnarray}
1=n_{1\uparrow}\geq n_{1\downarrow}=n_{2\uparrow}\geq n_{3\uparrow}=n_{2\downarrow}\geq n_{3\downarrow}=0.
\end{eqnarray}
The wave function now reads as
\bea\nonumber
\lefteqn{|S=1/2, S_z=1/2\rangle=}\hspace{1.2cm} \\
 & c_{121} |1^\uparrow2^\uparrow1'^\downarrow\rangle
+c_{122} |1^\uparrow2^\uparrow2'^\downarrow\rangle \nonumber\\
+e^{i\theta} & c_{122} |1^\uparrow3^\uparrow1'^\downarrow\rangle
+c_{132} |1^\uparrow3^\uparrow2'^\downarrow\rangle,
\label{eq:doublet2}
\eea
where $\theta$ is an arbitrary real number implying that the phase of the coefficient $c_{131}$ can be different from the phase of the coeffficient $c_{122}$. At this point, we need to remember that we are using natural orbitals for the single-particle orbitals, hence, the 1RDM has to be diagonal in those orbitals. Calculating the 1RDM of the above wave function we get the following constraint on the coefficients
\begin{equation}
c_{122}(e^{i\theta}c_{121}+c_{132})=0
\end{equation}
in order to avoid off-diagonal elements formed by the orbitals $2^\uparrow$ and $3^\uparrow$ or ${1'}^\downarrow$ and ${2'}^\downarrow$ . Consequently, either $c_{122}=0$ or $c_{132}=-e^{i\theta}c_{121}$. In the first case, the wave function would consist of only two determinants, $|1^\uparrow2^\uparrow1'^\downarrow\rangle$ and $|1^\uparrow3^\uparrow2'^\downarrow\rangle$. In the second case, we find four degenerate occupation numbers, i.e.\ $n_{1\downarrow}=n_{2\uparrow}=n_{3\uparrow}=n_{2\downarrow}=|c_{121}|^2+|c_{122}|^2$. As a consequence of the degenerate occupation numbers, a linear combination of two natural orbitals with the same spin and the same occupation number is again a natural orbital with that occupation. Choosing the linear combinations
\begin{eqnarray}\nonumber
\tilde{1}^\downarrow &=& \sqrt{2}[c_{121}1'^\downarrow+c_{122}2'^\downarrow]\\
\tilde{2}^\downarrow &=& \sqrt{2}e^{i\theta}[c_{122}1'^\downarrow-c_{121}2'^\downarrow],
\end{eqnarray}
we can rewrite the wave function as $[|1^\uparrow2^\uparrow\tilde{1}^\downarrow\rangle+e^{i\theta}|1^\uparrow3^\uparrow\tilde{2}^\downarrow\rangle]/\sqrt{2}$. As one can see, the four degenerate occupation numbers are all equal to 1/2. The inequality constraint is again satisfied as an equality in this case. We also note that, due to the degeneracy, these two cases could also be ordered as in iv), i.e.\ the case that we encounter numerically.

The cases i), iv), and v), apart from the $n_{1\uparrow}+n_{3\downarrow}=1$, have the following two additional equality GPCs
\begin{equation}
n_{1\downarrow}+n_{3\uparrow}=1, \: n_{2\uparrow}+n_{2\downarrow}=1.
\end{equation}
Note that the ordering iv) is the one we encounter numerically in our calculations. Acting on the wavefunction with these GPCs only the coefficients $c_{121}$, $c_{132}$, and $c_{233}$ can be non-zero. In the ordering i), $n_{1\downarrow}\geq n_{2\uparrow}$ requires that $c_{233}=0$ which means that $n_{3\downarrow}=0$, $n_{2\uparrow}=n_{1\downarrow}$, and $n_{3\uparrow}=n_{2\downarrow}$ . 
The inequality GPC
\begin{equation}
n_{3\uparrow}+n_{3\downarrow}-n_{2\downarrow}\geq 0
\end{equation}
is then satisfied as an equality. Due to the equalities for the occupation numbers, this case can also be ordered as in iv) and be regarded as a special case there. The wave function consists of the two Slater determinants $|1^\uparrow2^\uparrow1'^\downarrow\rangle$ and $|1^\uparrow3^\uparrow2'^\downarrow\rangle$. 
Again, the appearance of degenerate occupation numbers makes the removal of determinants due to the pinned GPCs questionable. However, enforcing the degeneracy $n_{2\uparrow}=n_{1\downarrow}$ at the beginning, the ordering i) can be seen as a special case of either ii) or iii). Using the same arguments as for these In case iv) the inequality reads
\begin{equation}
n_{2\downarrow}+n_{3\downarrow}-n_{3\uparrow}\geq 0
\end{equation}
is satisfied as an equality. Using the fact that $n_{2\downarrow}+n_{3\downarrow}=1-n_{1\downarrow}$, we see that the inequality reduces to one of the equality GPCs, thus is satisfied as equality.
For the ordering v) the inequality reads
\begin{equation}
n_{2\downarrow}+n_{3\downarrow}-n_{1\downarrow}\geq 0
\end{equation}
which implies that $n_{1\downarrow}\leq 0.5$, taking into account that the down occupation numbers sum up to 1. However, $n_{1\downarrow}$ being less or equal to $0.5$ seems to be quite unlikely in practice and explains why we never found such a case in our numerical results. Note that in all systems that we calculated the smallest $n_{1\downarrow}$ that we found is $0.98$. The wave function for both cases iv) and v) consists of three Slater determinants, namely 
\bea\nonumber
\hspace{2cm}|S=1/2, S_z=1/2\rangle=c_{121} |1^\uparrow2^\uparrow1'^\downarrow\rangle
\\\nonumber
+c_{132} |1^\uparrow3^\uparrow2'^\downarrow\rangle
+c_{233} |2^\uparrow3^\uparrow3'^\downarrow\rangle\\
\label{eq:doublet3}
\eea
the inequality constraint is pinned for case iv) which is the one that we typically encounter while it is not pinned in case v).

\item[b)] $n_{3\uparrow}$ is the smallest occupation number.

We can immediately conclude that the second smallest occupation number is $n_{3\downarrow}$ since, as we discussed before, the two smallest occupation numbers cannot belong both to spin up. The first GPC, $n_{1\uparrow}+n_{3\uparrow}=1$ implies that $n_{2\uparrow}=1$ since the spin-up occupations sum up to two. From the ordering of occupations it follows that $n_{1\uparrow}=1$ and consequently $n_{3\uparrow}=0$. 
There are then two options for the second largest occupation number, $n_{2\uparrow}$ or $n_{1\downarrow}$ with the second GPC being given by
\begin{eqnarray}
i) & n_{2\uparrow}+n_{3\downarrow}=1,\\
ii) & n_{1\downarrow}+n_{3\downarrow}=1,
\end{eqnarray}
respectively. For case i), it follows that $n_{3\downarrow}=0$ and only the coefficients $c_{121}$ and $c_{122}$ can be non-zero. The ordering is given by 
\begin{equation}
n_{1\uparrow}\geq n_{2\uparrow}\geq n_{1\downarrow}\geq n_{2\downarrow}\geq n_{2\downarrow}\geq n_{3\uparrow}
\end{equation}
and the inequality GPC reads $n_{3\downarrow}+n_{3\uparrow}-n_{2\downarrow}\geq 0$ which can only be satisfied if $n_{2\downarrow}=0$. Hence, the total wave function for this case is the single Slater determinant $|1^\uparrow2^\uparrow1'^\downarrow\rangle$. For case ii) we conclude from the ordering that $n_{1\downarrow}=1$ which only leaves the single Slater determinant $|1^\uparrow2^\uparrow1'^\downarrow\rangle$ to contribute to the wave function.
\end{itemize}
\end{itemize}
To conclude, the structure of the wave function in terms of natural orbitals in the 3-6 doublet case can contain maximally three Slater determinants, namely the $|1^\uparrow2^\uparrow1'^\downarrow\rangle$, $|1^\uparrow3^\uparrow2'^\downarrow\rangle$, and
$|2^\uparrow3^\uparrow3'^\downarrow\rangle$. The only possible orderings of the occupation numbers, which do not imply a single Slater determinant as the wave function, read as
\begin{eqnarray}
 &  n_{1\uparrow}\geq n_{2\uparrow}\geq n_{1\downarrow}\geq n_{3\uparrow}\geq n_{2\downarrow}\geq n_{3\downarrow},\\
 &  n_{1\uparrow}\geq n_{2\uparrow}\geq n_{3\uparrow}\geq n_{1\downarrow}\geq n_{2\downarrow}\geq n_{3\downarrow}
\end{eqnarray}
as all the others can be considered as special cases of the above. The first one is the one we always found numerically, while the second one requires that $n_{1\downarrow}$ is smaller than $0.5$. The inequality constraint is pinned for the first ordering while for the second one it is not pinned.
The list of equality GPCs in the 3-6 case, when written in terms of spin occupation numbers, is unique,
\begin{align}
\nonumber
n_{1\uparrow}+n_{3\downarrow}&=1,& 
n_{2\uparrow}+n_{2\downarrow}=1,\\
n_{1\downarrow}+n_{3\uparrow}&=1.
\end{align}
as the two possible orderings differ only in the interchange between the third and the fourth occupation numbers in descending order.

\subsection{$N=3, M=8$}
\label{app:3-8GPC}
Contrary to the 3-6 case, where the ordering of the spin occupation numbers that we encounter numerically is unique, for the 3-8 case we found three different orderings in our CASSCF calculations in the space of dimension 8. Theoretically, of course, there are many more possible orderings for this case. For the three cases which we found numerically, we provide the full list of generalized Pauli constraints written using spin indexed occupation numbers. 

\begin{itemize}
\item [Case 1:] $n_{1\uparrow} \geq n_{2\uparrow} \geq n_{1\downarrow} \geq n_{3\uparrow} \geq n_{2\downarrow} \geq n_{3\downarrow} \geq n_{4\uparrow} \geq n_{4\downarrow}$
\hspace*{-1.2cm}
\begin{tabular}{r|l}
\multicolumn{2}{c}{ }\\
\label{tab:case1}
\# & Condition\\ \hline
1. & $2-n_{1\uparrow}-n_{2\uparrow}-n _{3\uparrow}-n_{4\uparrow} \geq 0$\\
2. & $2-n_{1\uparrow}-n_{2\uparrow}-n_{2\downarrow}-n_{3\downarrow} \geq 0$\\
3. & $2-n_{2\uparrow}-n_{1\downarrow}-n_{3\uparrow}-n_{2\downarrow}\geq 0$\\
4. & $2-n_{1\uparrow}-n_{1\downarrow}-n_{3\uparrow}-n_{3\downarrow} \geq 0$\\ \hline
5. & $1-n_{1\uparrow}-n_{2\uparrow}+n_{1\downarrow}\geq 0$\\
6. & $1-n_{2\uparrow}-n_{2\downarrow}+n_{4\uparrow} \geq 0$\\
7. & $1-n_{1\uparrow}-n_{3\downarrow}+n_{4\uparrow} \geq 0$\\
8. & $1-n_{2\uparrow}-n_{3\uparrow}+n_{3\downarrow} \geq 0$\\
9. & $1-n_{1\uparrow}-n_{3\uparrow}+n_{2\downarrow}\geq 0$\\
10. & $1-n_{1\downarrow}-n_{3\uparrow}+n_{4\uparrow} \geq 0$\\ \hline
11. & $1-n_{1\uparrow}-n_{4\downarrow}\geq 0$\\ \hline
12. & $0-n_{2\uparrow}+n_{1\downarrow}+n_{3\downarrow}+n_{4\uparrow}\geq 0$\\
13. & $0-n_{3\uparrow}+n_{2\downarrow}+n_{3\downarrow}+n_{4\uparrow}\geq 0$\\
14. & $0-n_{1\uparrow}+n_{1\downarrow}+n_{2\downarrow}+n_{4\uparrow} \geq 0$\\ \hline
15. & $2-n_{2\uparrow}-n_{1\downarrow}-2n_{3\uparrow}+n_{2\downarrow}+n_{4\uparrow}-n_{4\downarrow}\geq 0$\\
16. & $2-n_{1\uparrow}-n_{1\downarrow}-2n_{3\uparrow}+n_{2\downarrow}+n_{3\downarrow}-n_{4\downarrow}\geq 0$\\
17. & $2-n_{1\uparrow}-2n_{2\uparrow}+n_{1\downarrow}-n_{3\uparrow}+n_{2\downarrow}-n_{4\downarrow}\geq 0$\\
18. & $ 2-n_{1\uparrow}-2n_{2\uparrow}+n_{1\downarrow}-n_{2\downarrow}+n_{3\downarrow}-n_{4\downarrow} \geq 0$\\ \hline
19. & $0-n_{1\uparrow}-n_{2\uparrow}+2n_{1\downarrow}+n_{3\uparrow}+n_{2\downarrow}\geq 0$\\
20. & $0-n_{1\uparrow}+n_{2\uparrow}+n_{1\downarrow}-n_{3\downarrow}+2n_{4\uparrow} \geq 0$\\ \hline
21. & $0-n_{1\uparrow}+n_{1\downarrow}+n_{3\uparrow}+n_{2\downarrow}-n_{4\downarrow}\geq 0$\\
22. & $0-n_{1\uparrow}+n_{2\uparrow}+n_{1\downarrow}+n_{4\uparrow}-n_{4\downarrow} \geq 0$\\ \hline
23. & $1-2n_{1\uparrow}+n_{2\uparrow}-n_{3\uparrow}+2n_{2\downarrow}+n_{3\downarrow}-n_{4\downarrow} \geq 0$\\
24. & $1-n_{1\downarrow}-2n_{3\uparrow}+2n_{2\downarrow}+n_{3\downarrow}+n_{4\uparrow} -n_{4\downarrow} \geq 0$\\
25. & $1-2n_{1\uparrow}+n_{2\uparrow}+n_{3\uparrow}-n_{3\downarrow}+2n_{4\uparrow}-n_{4\downarrow}  \geq 0$\\
26. & $1-2n_{1\uparrow}-n_{2\uparrow}+2n_{1\downarrow}+n_{3\uparrow}+n_{3\downarrow}-n_{4\downarrow}  \geq 0$\\
27. & $1-n_{1\uparrow}-2n_{2\uparrow}+2n_{1\downarrow}+n_{2\downarrow}+n_{3\downarrow}-n_{4\downarrow}  \geq 0$\\ \hline
28. & $0-2n_{1\uparrow}+2n_{2\uparrow}+n_{1\downarrow}+n_{3\uparrow}-n_{3\downarrow}+3n_{4\uparrow}-n_{4\downarrow}  \geq 0$\\
29. & $0+n_{1\uparrow}-n_{1\downarrow}-2n_{3\uparrow}+3n_{2\downarrow}+2n_{3\downarrow}+n_{4\uparrow}-n_{4\downarrow}  \geq 0$\\
30. & $0-2n_{1\uparrow}-n_{2\uparrow}+3n_{1\downarrow}+2n_{3\uparrow}+n_{2\downarrow}+n_{3\downarrow}-n_{4\downarrow}  \geq 0$\\
31. & $0-n_{1\uparrow}-2n_{2\uparrow}+3n_{1\downarrow}+n_{3\uparrow}+2n_{2\downarrow}+n_{3\downarrow}-n_{4\downarrow}  \geq 0$
\end{tabular}
\newpage
\item[Case 2:] $n_{1\uparrow} \geq n_{2\uparrow} \geq n_{1\downarrow} \geq n_{3\uparrow} \geq n_{2\downarrow} \geq n_{4\uparrow} \geq n_{3\downarrow} \geq n_{4\downarrow}$
\hspace*{-1.2cm}
\begin{tabular}{r|l}
\multicolumn{2}{c}{ }\\
\# & Condition\\ \hline
1. & $2-n_{1\uparrow}-n_{2\uparrow}-n_{3\uparrow}-n_{3\downarrow}\geq 0$\\
2. & $2-n_{1\uparrow}-n_{2\uparrow}-n_{2\downarrow}-n_{4\uparrow} \geq 0$\\
3. & $2-n_{2\uparrow}-n_{1\downarrow}-n_{3\uparrow}-n_{2\downarrow}\geq 0$\\
4. & $2-n_{1\uparrow}-n_{1\downarrow}-n_{3\uparrow}-n_{4\uparrow} \geq 0$\\ \hline
5. & $1-n_{1\uparrow}-n_{2\uparrow}+n_{1\downarrow}\geq 0$\\
6. & $1-n_{2\uparrow}-n_{2\downarrow}+n_{3\downarrow} \geq 0$\\
7. & $1-n_{1\uparrow}-n_{4\uparrow}+n_{3\downarrow} \geq 0$\\
8. & $1-n_{2\uparrow}-n_{3\uparrow}+n_{4\uparrow} \geq 0$\\
9. & $1-n_{1\uparrow}-n_{3\uparrow}+n_{2\downarrow}\geq 0$\\
10. & $1-n_{1 \downarrow}-n_{3\uparrow}+n_{3\downarrow} \geq 0$\\ \hline
11. & $1-n_{1\uparrow}-n_{4\downarrow}\geq 0$\\ \hline
12. & $0-n_{2\uparrow}+n_{1\downarrow}+n_{3\downarrow}+n_{4\uparrow}\geq 0$\\
13. & $0-n_{3\uparrow}+n_{2\downarrow}+n_{3\downarrow}+n_{4\uparrow}\geq 0$\\
14. & $0-n_{1\uparrow}+n_{1\downarrow}+n_{2\downarrow}+n_{3\downarrow} \geq 0$\\ \hline
15. & $2-n_{2\uparrow}-n_{1\downarrow}-2n_{3\uparrow}+n_{2\downarrow}+n_{3\downarrow}-n_{4\downarrow}\geq 0$\\
16. & $2-n_{1\uparrow}-n_{1\downarrow}-2n_{3\uparrow}+n_{2\downarrow}+n_{4\uparrow}-n_{4\downarrow}\geq 0$\\
17. & $2-n_{1\uparrow}-2n_{2\uparrow}+n_{1\downarrow}-n_{3\uparrow}+n_{2\downarrow}-n_{4\downarrow}\geq 0$\\
18. & $2-n_{1\uparrow}-2n_{2\uparrow}+n_{1\downarrow}-n_{2\downarrow}+n_{4\uparrow}-n_{4\downarrow}\geq 0$\\ \hline
19. & $0-n_{1\uparrow}-n_{2\uparrow}+2n_{1\downarrow}+n_{3\uparrow}+n_{2\downarrow}\geq 0$\\
20. & $0-n_{1\uparrow}+n_{2\uparrow}+n_{1\downarrow}-n_{4\uparrow}+2n_{3\downarrow} \geq 0$\\ \hline
21. & $0-n_{1\uparrow}+n_{1\downarrow}+n_{3\uparrow}+n_{2\downarrow}-n_{4\downarrow}\geq 0$\\
22. & $0-n_{1\uparrow}+n_{2\uparrow}+n_{1\downarrow}+n_{3\downarrow}-n_{4\downarrow} \geq 0$\\ \hline
23. & $1-2n_{1\uparrow}+n_{2\uparrow}-n_{3\uparrow}+2n_{2\downarrow}+n_{4\uparrow}-n_{4\downarrow} \geq 0$\\
24. & $1-n_{1\downarrow}-2n_{3\uparrow}+2n_{2\downarrow}+n_{3\downarrow}+n_{4\uparrow} -n_{4\downarrow}\geq 0$\\
25. & $1-2n_{1\uparrow}+n_{2\uparrow}+n_{3\uparrow}-n_{4\uparrow}+2n_{3\downarrow}-n_{4\downarrow}  \geq 0$\\
26. & $1-2n_{1\uparrow}-n_{2\uparrow}+2n_{1\downarrow}+n_{3\uparrow}+n_{4\uparrow}-n_{4\downarrow}  \geq 0$\\
27. & $1-n_{1\uparrow}-2n_{2\uparrow}+2n_{1\downarrow}+n_{2\downarrow}+n_{4\uparrow}-n_{4\downarrow}  \geq 0$\\ \hline
28. & $0-2n_{1\uparrow}+n_{2\uparrow}+n_{1\downarrow}+n_{3\uparrow}-n_{4\uparrow}+3n_{3\downarrow}-n_{4\downarrow}  \geq 0$\\
29. & $0+n_{1\uparrow}-n_{1\downarrow}-2n_{3\uparrow}+3n_{2\downarrow}+2n_{4\uparrow}+n_{3\downarrow}-n_{4\downarrow}  \geq 0$\\
30. & $0-2n_{1\uparrow}-n_{2\uparrow}+3n_{1\downarrow}+2n_{3\uparrow}+n_{2\downarrow}+n_{4\uparrow}-n_{4\downarrow}  \geq 0$\\
31. & $0-n_{1\uparrow}-2n_{2\uparrow}+3n_{1\downarrow}+n_{3\uparrow}+2n_{2\downarrow}+n_{4\uparrow}-n_{4\downarrow}  \geq 0$
\end{tabular}
\newpage
\item[Case 3:] $n_{1\uparrow} \geq n_{1\downarrow} \geq n_{2\uparrow} \geq n_{3\uparrow} \geq n_{2\downarrow} \geq n_{4\uparrow} \geq n_{3\downarrow} \geq n_{4\downarrow}$
\hspace*{-1.2cm}
\begin{tabular}{r|l}
\multicolumn{2}{c}{ }\\
\# & Condition\\ \hline
1. & $2-n_{1\uparrow}-n_{1\downarrow}-n_{3\uparrow}-n_{3\downarrow} \geq 0$\\
2. & $2-n_{1\uparrow}-n_{1\downarrow}-n_{2\downarrow}-n_{4\uparrow} \geq 0$\\
3. & $2-n_{1\downarrow}-n_{2\uparrow}-n_{3\uparrow}-n_{2\downarrow}\geq 0$\\
4. & $2-n_{1\uparrow}-n_{2\uparrow}-n_{3\uparrow}-n_{4\uparrow} \geq 0$\\ \hline
5. & $1-n_{1\uparrow}-n_{1\downarrow}+n_{2\uparrow}\geq 0$\\
6. & $1-n_{1\downarrow}-n_{2\downarrow}+n_{3\downarrow} \geq 0$\\
7. & $1-n_{1\uparrow}-n_{4\uparrow}+n_{3\downarrow} \geq 0$\\
8. & $1-n_{1\downarrow}-n_{3\uparrow}+n_{4\uparrow} \geq 0$\\
9. & $1-n_{1\uparrow}-n_{3\uparrow}+n_{2\downarrow}\geq 0$\\
10. & $1-n_{2\uparrow}-n_{3\uparrow}+n_{3\downarrow} \geq 0$\\ \hline
11. & $1-n_{1\uparrow}-n_{4\downarrow}\geq 0$\\ \hline
12. & $0-n_{1\downarrow}+n_{2\uparrow}+n_{3\downarrow}+n_{4\uparrow}\geq 0$\\
13. & $0-n_{3\uparrow}+n_{2\downarrow}+n_{3\downarrow}+n_{4\uparrow}\geq 0$\\
14. & $0-n_{1\uparrow}+n_{2\uparrow}+n_{2\downarrow}+n_{3\downarrow}\geq 0$\\ \hline
15. & $2-n_{1\downarrow}-n_{ 2\uparrow}-2n_{3\uparrow}+n_{2\downarrow}+n_{3\downarrow}-n_{4\downarrow}\geq 0$\\
16. & $2-n_{1\uparrow}-n_{2\uparrow}-2n_{3\uparrow}+n_{2\downarrow}+n_{4\uparrow}-n_{4\downarrow}\geq 0$\\
17. & $2-n_{1\uparrow}-2n_{1\downarrow}+n_{2\uparrow}-n_{3\uparrow}+n_{2\downarrow}-n_{4\downarrow}\geq 0$\\
18. & $2-n_{1\uparrow}-2n_{1\downarrow}+n_{2\uparrow}-n_{2\downarrow}+n_{4\uparrow}-n_{4\downarrow}\geq 0$\\ \hline
19. & $0-n_{1\uparrow}-n_{1\downarrow}+2n_{2\uparrow}+n_{3\uparrow}+n_{2\downarrow}\geq 0$\\
20. & $0-n_{1\uparrow}+n_{1\downarrow}+n_{2\uparrow}-n_{4\uparrow}+2n_{3\downarrow} \geq 0$\\ \hline
21. & $0-n_{1\uparrow}+n_{2\uparrow}+n_{3\uparrow}+n_{2\downarrow}-n_{4\downarrow}\geq 0$\\
22. & $0-n_{1\uparrow}+n_{1\downarrow}+n_{2\uparrow}+n_{3\downarrow}-n_{4\downarrow} \geq 0$\\ \hline
23. & $1-2n_{1\uparrow}+n_{1\downarrow}-n_{3\uparrow}+2n_{2\downarrow}+n_{4\uparrow}-n_{4\downarrow} \geq 0$\\
24. & $1-n_{2\uparrow}-2n_{3\uparrow}+2n_{2\downarrow}+n_{3\downarrow}+n_{4\uparrow} -n_{4\downarrow}\geq 0$\\
25. & $1-2n_{1\uparrow}+n_{1\downarrow}+n_{3\uparrow}-n_{4\uparrow}+2n_{3\downarrow}-n_{4\downarrow}  \geq 0$\\
26. & $1-2n_{1\uparrow}-n_{1\downarrow}+2n_{2\uparrow}+n_{3\uparrow}+n_{4\uparrow}-n_{4\downarrow}  \geq 0$\\
27. & $1-n_{1\uparrow}-2n_{1\downarrow}+2n_{2\uparrow}+n_{2\downarrow}+n_{4\uparrow}-n_{4\downarrow}  \geq 0$\\ \hline
28. & $0-2n_{1\uparrow}+2n_{1\downarrow}+n_{2\uparrow}+n_{3\uparrow}-n_{4\uparrow}+3n_{3\downarrow}-n_{4\downarrow} \geq 0$\\
29. & $0+n_{1\uparrow}-n_{2\uparrow}-2n_{3\uparrow}+3n_{2\downarrow}+2n_{4\uparrow}+n_{3\downarrow}-n_{4\downarrow}  \geq 0$\\
30. & $0-2n_{1\uparrow}-n_{1\downarrow}+3n_{2\uparrow}+2n_{3\uparrow}+n_{2\downarrow}+n_{4\uparrow}-n_{4\downarrow} \geq 0$\\
31. & $0-n_{1\uparrow}-2n_{1\downarrow}+3n_{2\uparrow}+n_{3\uparrow}+2n_{2\downarrow}+n_{4\uparrow}-n_{4\downarrow}\geq 0$
\end{tabular}

\end{itemize}

\section{Relations between the the expansion coefficients of CI and the spin natural orbitals}\label{app:eigenstate_cond}

In this Appendix, we discuss the conditions that need to be satisfied in order for the CI expansion in the 3-6 case to form a spin doublet eigenstate.
As discussed in section \ref{sec:spin_dependence}, in order to describe spin eigenstates of the many electron wavefunction, the spatial parts of the natural orbitals are in general different for the up and down spin channels. However, they still have to span the same space. This conclusion is derived from the following example. For three electrons in $M=6$ natural orbitals we choose three spatial orbitals for each spin channel. Let as assume that the spatial parts of the down natural orbitals $j'$ are linear combinations of two up ones $k$, plus one basis function $\chi$ that does not belong to the space spanned by the up orbitals

\begin{equation}
j' = \sum_{k=1}^2 (k|j')k + (\chi|j') \chi,
\label{eq:down_orb}
\end{equation}

Using this expansion in the wave function with $S=1/2$, see Eq.\ (\ref{eq:psi_spin_half}), we find

\begin{eqnarray}
\nonumber
\lefteqn{|S=1/2, S_z=1/2\rangle=}\hspace{1cm}\\
\nonumber
&& c_{121} (1|1')|1^\uparrow 2^\uparrow 1^\downarrow\rangle +c_{121}(2|1')|1^\uparrow2^\uparrow2^\downarrow\rangle\\
\nonumber
&+&c_{132} (1|2')|1^\uparrow3^\uparrow 1^\downarrow\rangle
+ c_{132} (2|2')|1^\uparrow3^\uparrow 2^\downarrow\rangle\\
\nonumber
&+&c_{233} (1|3')| 2^\uparrow 3^\uparrow 1^\downarrow\rangle +c_{233}(2|3')|2^\uparrow 3^\uparrow 2^\downarrow\rangle
\\
\nonumber
&+& c_{121}(\chi|1')|1^\uparrow 2^\uparrow\chi^\downarrow\rangle+c_{132}(\chi|2')|1^\uparrow 3^\uparrow\chi^\downarrow\rangle\\
\label{eq:psi_chi}
&+&c_{233}(\chi|3')|2^\uparrow 3^\uparrow \chi^\downarrow\rangle
\end{eqnarray}
Since our state has maximum $S_z$ acting with the operator $\mathbf{S^{+}}$ gives zero.
In doing so, we find three linearly independent Slater determinants namely the ones that contain the orbitals $\chi$:
$|1^\uparrow 2^\uparrow \chi^\uparrow\rangle$,
$|1^\uparrow 3^\uparrow \chi^\uparrow\rangle$,
$|2^\uparrow 3^\uparrow \chi^\uparrow\rangle$
which can only be zero if
$(\chi|1')=(\chi|2')=(\chi|3')=0$, which is contrary to our assumption \ref{eq:down_orb}.

As a result, one can expand one set of natural spin orbitals in the orbitals of the opposite spin channel. Such an expansion reads as
\begin{equation}\label{eq:expansion}
j^\sigma = \sum_{k=1}^{M/2} (k^{\bar{\sigma}} |j^\sigma) k^{\bar{\sigma}},
\end{equation}
where $\bar{\sigma}$ denotes the opposite spin of $\sigma$. Consequently, one can use this expansion and transform the CI expansion in Slater determinants of natural orbitals to an expansion in Slater determinants of spin-independent orbitals. In the following, we again use the example of three electrons and $M=6$.
 
As discussed in section \ref{sec:spin_dependence}, a CI expansion of a spin eigenstate with $S=1/2$ in the basis of the natural orbitals contains only three Slater determinants, namely
\begin{equation}
|\Psi\rangle=c_{121}|1^\uparrow 2^\uparrow 1'^{\downarrow}\rangle+c_{132}|1^\uparrow 3^\uparrow 2'^{\downarrow}\rangle +c_{233}|2^\uparrow 3^\uparrow 3'^{\downarrow}\rangle.
\end{equation}
Using Eq.\ (\ref{eq:expansion}) to expand the orbitals of the down channel, we obtain for $\Psi$
\begin{eqnarray}\label{eq:psi_expanded}
|\Psi\rangle=\sum_{k=1}^3 \left( c_{121} (k|1')|1^\uparrow 2^\uparrow k^\downarrow\rangle
+c_{132} (k|2')|1^\uparrow 3^\uparrow k^\downarrow \rangle \right . \nonumber\\
\left . +c_{233} (k|3')|2^\uparrow 3^\uparrow k^\downarrow\rangle \right).
\end{eqnarray}
 As the orbitals in Eq.\ (\ref{eq:psi_expanded}) have the same spatial parts in both spin channels, any Slater determinant with a doubly occupied orbital is a spin eigenstate with $S=1/2$. The only contributions which are not eigenstates by themselves are therefore given by
\begin{equation}
c_{121} (3|1')|1^\uparrow 2^\uparrow 3^\downarrow \rangle
+c_{132} (2|2') |1^\uparrow 3^\uparrow 2^\downarrow\rangle
+c_{233} (1|3') |2^\uparrow 3^\uparrow 1^\downarrow\rangle.
\label{spin_cont}
\end{equation}
In order for this part of the wavefunction to form a spin eigenstate with $S=S_z=1/2$ the following condition should hold
\begin{equation}
c_{123} (3|1')-c_{132} (2|2')+c_{233} (1|3') =0,
\end{equation}
which can be derived by acting on Eq.~(\ref{spin_cont}) with $\mathbf{S^{+}}$ and requiring that it gives zero.
As one can see, the overlaps $(k|j')$ between the spatial parts of the spin up and the spin down orbitals enter this equation. Hence, the condition on the CI coefficients so that the corresponding wave function is a spin eigenstate, depends on the specific relation between the natural orbitals in the two spin channels.

We emphasize that the expansion Eq.~(\ref{eq:expansion}) can be done for any $M$. However, due to the increasing number of orbitals in Eq.\ (\ref{eq:expansion}) and an increasing number of Slater determinants in the CI expansion of $\Psi$, the relations between the CI coefficients become more complicated. Also, since only three orbitals enter the Slater determinants, one obtains more than one relation for $\frac{M}{2}>3$. For example, for $M=8$ one finds four different relations, one for each set of determinants that have a specific orbital missing.

\section*{Acknowledgments}
We thank Carlos L. Benavides-Riveros for very helpful discussions.
NNL acknowledges support by the project ``Advanced Materials and Devices'' (MIS 5002409) which is implemented under the ``Action for the Strategic Development on the Research and Technological Sector'', funded by the Operational Program ``Competitiveness, Entrepreneurship and Innovation'' (NSRF 2014-2020) and co-financed by Greece and the European Union (European Regional Development Fund). 
NH acknowledges support from an Emmy-Noether grant from Deutsche Forschungsgemeinschaft.

\bibliographystyle{apsrev}
\bibliography{spin_gpc}

\end{document}